\documentclass{pasj01}
\let\leftroot\undefined
\let\uproot\undefined

\usepackage{amsmath}
\usepackage[switch]{lineno}
\Received{$\langle$reception date$\rangle$}
\Accepted{$\langle$acception date$\rangle$}
\Published{$\langle$publication date$\rangle$}

\begin{document}

\title{SN 2023dbc in M108: Optical and Near-Infrared Observations of a Highly-Obscured, Moderately Energetic Stripped-Envelope Supernova}
\author{Masayuki Yamanaka$^{1}$, Takahiro Nagayama$^{2}$, Akari Kumano$^{3}$, Devendra Kumar Sahu$^{4}$, Avinash Singh$^{5}$, Hrishav Das$^{4}$, and G. C. Anupama$^{4}$}

\altaffiltext{1}{Amanogawa Galaxy Astronomy Research Center (AGARC), Graduate School of Science and Engineering, Kagoshima University, 1-21-35 Korimoto, Kagoshima, Kagoshima 890-0065, Japan}
\email{yamanaka@sci.kagoshima-u.ac.jp}

\altaffiltext{2}{Graduate School of Science and Engineering, Kagoshima University, 1-21-35 Korimoto, Kagoshima, Kagoshima 890-0065, Japan}
\altaffiltext{3}{Department of Physics and Astronomy, Faculty of Science, Kagoshima University, 1-21-35 Korimoto,
Kagoshima, Kagoshima 890-0065, Japan}
\altaffiltext{4}{Indian Institute of Astrophysics, Koramangala 2nd Block, Bangalore 560034, India}
\altaffiltext{5}{Oskar Klein Centre, Department of Astronomy, Stockholm University, AlbaNova, SE-106 91 Stockholm, Sweden}


\maketitle

\begin{abstract}
We present near-infrared (NIR) and optical observations of the highly reddened and moderately energetic Type Ib supernova (SN) 2023dbc, {\bf covering a period from} 2 to 136 days after the explosion. By comparing its color {\bf evolution}, specifically in $r-JHK_{\mathrm{s}}$ and $i-JHK_{\mathrm{s}}$, with those of broad-lined Type Ic (Ic-BL) and Type IIb SNe, we estimate a significant extinction of $A_{V}=4.1\pm0.1$\,mag toward the SN. The extinction-corrected peak absolute magnitudes are $M_{J} = -16.8\pm0.2$\,mag, $M_{H} = -16.8\pm0.2$\,mag, and $M_{K_{\mathrm{s}}} = -17.0\pm0.2$\,mag. The SN {\bf exhibited} an $r$-band rise time of 14.9 days. The spectra {\bf display} broad features {\bf indicative of} high expansion velocities; the He~{\sc i} line velocity was measured at $16,000\,\mathrm{km\,s^{-1}}$ at $t=-4$\,d. Its spectral profile is broader than {\bf those} of typical moderately energetic Type Ib SNe, {\bf yet narrower than those of Type Ic-BL SNe, placing it in an intermediate category}. Based on the light-curve timescale and velocity, we estimate {\bf a} kinetic energy of $E_k = (4.1\pm0.7) \times 10^{51}$\,erg, {\bf an} ejecta mass of $M_{\mathrm{ej}} = 2.3\pm0.7\,M_{\odot}$, and a radioactive $^{56}\mathrm{Ni}$ mass of $(3.8\pm0.1) \times 10^{-2}\,M_{\odot}$. {\bf An} analysis using a two-component model suggests a steep density profile in the outer layer {\bf contrasted with} a dense inner core, {\bf which implies} ejecta asphericity. The low $^{56}\mathrm{Ni}$ mass is consistent with a partial fallback scenario. We conclude that SN 2023dbc originated from an aspherical explosion with partial core fallback, {\bf arising} from a progenitor ($M_{\mathrm{ini}} \simeq 15\,M_{\odot}$) that had retained {\bf its} helium envelope {\bf within} a binary system.
\end{abstract}

\section{Introduction}
Type Ib/c supernovae (SNe Ib/c) are widely known to be the core-collapse explosions of massive stars that have shed their hydrogen envelopes \citep{Nomoto1994,Woosley1995}. Specifically, the spectra of SNe Ib {\bf exhibit} helium lines, whereas SNe Ic {\bf lack such features} \citep{Filippenko1997}. However, {\bf because their} progenitors are relatively compact compared to typical supergiants and are often {\bf located} in crowded regions of {\bf their} host {\bf galaxies} \citep{Anderson2008,Hakobyan2009,Aramyan2016}, {\bf direct detections} of progenitor stars in pre-explosion images {\bf are} extremely challenging \citep{Cao2013,Groh2013,Kilpatrick2018,Zhao2025}. Consequently, the origin of SNe Ib/c remains {\bf a} highly {\bf debated} and unresolved {\bf issue in stellar astrophysics}.

Light curve and spectral analyses of SNe~Ib/c {\bf have revealed} low ejecta masses and typical kinetic energies \citep{Cano2013,Prentice2016,Lyman2016}. These results support the {\bf hypothesis} that the progenitors may {\bf originate} from low-mass stripped-envelope stars, {\bf rather} than from very massive stars \citep{Hachinger2012,Dessart2020}. Such low-mass progenitors are {\bf well} reproduced by binary interaction scenarios \citep{Yoon2010,Dessart2012,Tauris2015}. On the other hand, some highly energetic events are often associated with intense high-energy emission {\bf such as} gamma-ray {\bf bursts} \citep{Galama1998,Stanek2003,Malesani2004}. The progenitors of these events are {\bf suggested} to be Wolf--Rayet stars that evolved from massive stars with initial masses $> 25\,M_{\odot}$ \citep{Iwamoto1998,Iwamoto2000,Mazzali2003}. The event rate of SNe~Ib/c obtained {\bf from} volume-limited {\bf samples} is approximately {\bf one-third} of that of SNe~II \citep{Li2011,Graur2017,Ma2025}. Therefore, {\bf a larger number of} well-observed nearby samples {\bf is essential} for a comprehensive understanding.

{\bf Given the crowded environments of their explosion sites, some SNe~Ib/c are discovered within the spiral arms of their host galaxies. Consequently, they are often affected by significant extinction due to interstellar dust \citep{Kankare2014a,Kankare2014b,Jencson2017,Jencson2018,Kool2018,Kankare2021}. The intrinsic SN event rate in dusty environments remains uncertain \citep{Mattila2012}, and investigating such highly obscured events 
would enhance our understanding of the properties of these phenomena. 
Although follow-up observations of these highly reddened SNe are challenging at optical wavelengths, NIR observations are less affected because scattering and absorption are much weaker. However, the number of facilities capable of such NIR monitoring has been limited.}
 
SN~2023dbc was discovered at $r = 19.5$\,mag in {\bf a} spiral arm of the galaxy M108 on 2023 March 13 {\bf at} 06:53:43 UT \citep{Ho2023} by the Zwicky Transient Facility (ZTF; \cite{Bellm2019}). {\bf Remarkably}, the last non-detection limit was recorded at $r = 20.3$\,mag {\bf only} 40 minutes prior, on 2023 March 13 {\bf at} 06:16:42 UT \citep{Ho2023}. {\bf Subsequent} independent detection was provided by {\bf the} Asteroid Terrestrial-impact Last Alert System (ATLAS; \cite{Tonry2018}), {\bf which reported} $o = 17.5$\,mag on 2023 March 16 {\bf at} 10:54:10 UT. The corresponding non-detection limit was $o = 19.2$\,mag on 2023 March 12 {\bf at} 11:34:41 UT, approximately eighteen hours before the ZTF discovery. {\bf Additionally, we obtained forced photometry data from the ATLAS server} \citep{KWSmith2020,Shingles2021} {\bf to further constrain the early-phase light curve}.

Following the discovery, spectroscopic identification was {\bf carried out}. The object was initially classified as a young Type~II SN on March 16 {\bf based on a spectrum obtained with the} Lijiang-2.4m/YFOSC \citep{Li2023}. However, a {\bf subsequent} spectrum obtained {\bf with} Keck-I/LRIS \citep{Perley2023} identified the object {\bf as} a highly reddened Type~Ic SN {\bf in its early phases}. {\bf In this study, through our detailed multi-band light curve and spectroscopic analyses, we finally identify SN~2023dbc as a Type~Ib SN, characterized by the presence of prominent He~{\sc i} lines.}

The {\bf remainder of this paper is organized as follows}. In \S 2, we describe the observations and data reduction {\bf procedures} in detail. {\bf Section 3 presents the results of our photometric and spectroscopic analyses}, {\bf while} we discuss the {\bf derived} ejecta and progenitor properties in \S 4. Finally, we summarize our findings {\bf and conclude the study} in \S 5.

\begin{figure}
    \includegraphics[width=0.48\textwidth]{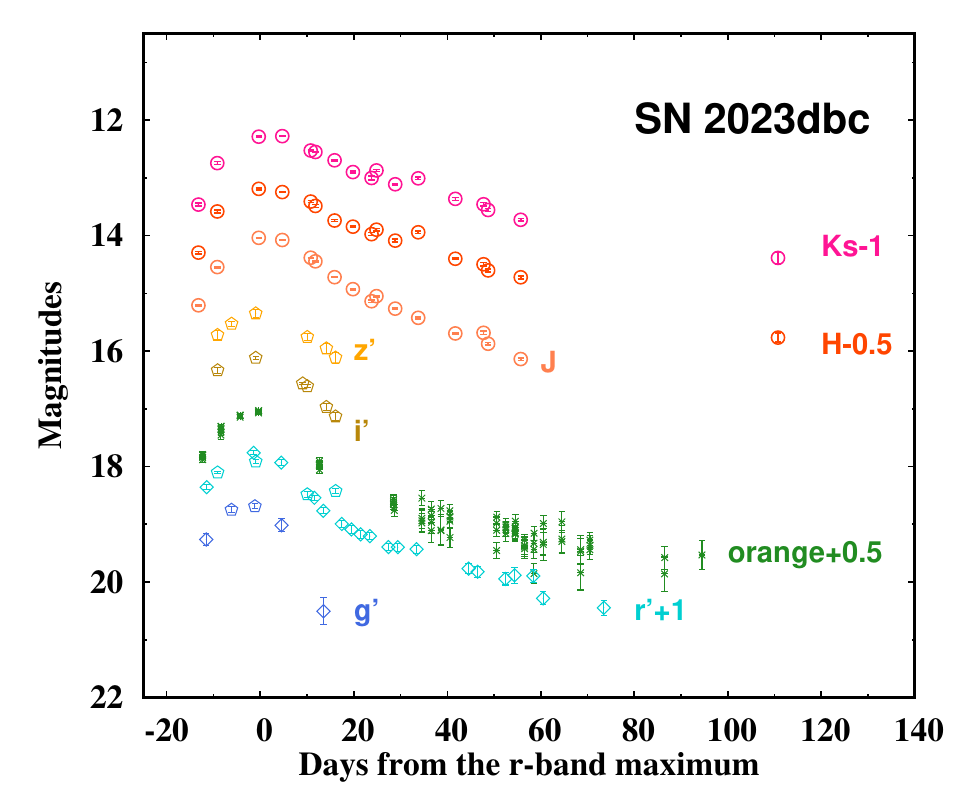}
     \caption{Optical and NIR $g'$, $r'$, $o$, $i$, $z$, $J$, $H$, and $K_{\rm s}$-band light curves of SN~2023dbc. 
{\bf Note that these} light curves {\bf have not been corrected for Galactic or host-galaxy extinction}. 
For clarity, {\bf the data are plotted} with {\bf arbitrary} vertical {\bf offsets}. 
The $g'$, $r'$, $o$, $i$, and $z$-band {\bf magnitudes are presented} in the AB system, {\bf whereas the} $J$, $H$, and $K_{\rm s}$ {\bf magnitudes are in the} Vega system. 
The $r$-band maximum date {\bf is adopted} as the reference {\bf epoch ($t = 0$)}. 
Alt text: The horizontal axis represents the phase in days relative to the $r$-band maximum light.}
\end{figure}

\begin{figure}
    \includegraphics[width=0.45\textwidth]{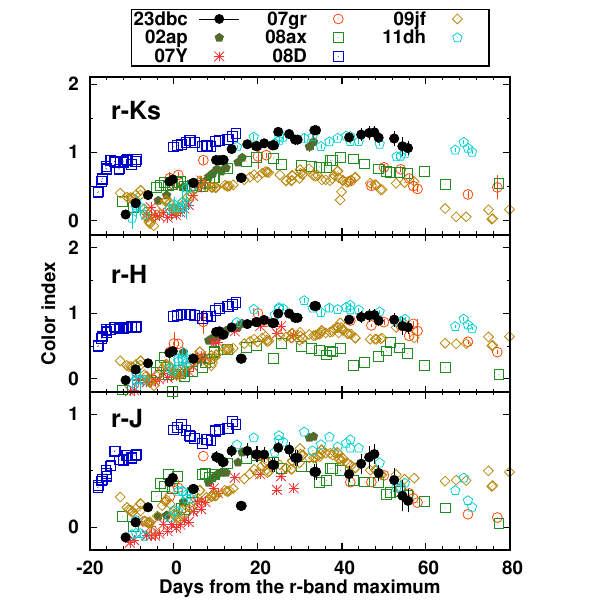} 
  \caption{{\bf Extinction-corrected} $r-K_{\mathrm{s}}$, $r-H$, and $r-J$ color {\bf evolutions} of SN~2023dbc. These are compared with those of SNe~2002ap \citep{Yoshii2003,Tomita2006}, 2007Y \citep{Stritzinger2009}, 2007gr \citep{Hunter2009}, 2008D \citep{Modjaz2009,Bianco2014}, 2008ax \citep{Pastorello2008,Bianco2014}, 2009jf \citep{Sahu2011,Valenti2011,Bianco2014}, and 2011dh \citep{Sahu2013,Ergon2014,Marion2014}. {\bf Note that} $R-K_{\mathrm{s}}$, $R-H$, and $R-J$ {\bf are plotted for the comparison}. The color evolution of SN~2011dh {\bf were adopted} as a template to derive the color excesses {\bf for SN~2023dbc}: $E(r-K_{\mathrm{s}})=2.9$, $E(r-H)=2.7$, and $E(r-J)=2.3${\bf . These values} were {\bf applied} to correct the {\bf observed} colors of SN~2023dbc. Alt text: Three line graphs showing color evolution. The horizontal axis represents the phase in days relative to the maximum light.}
\end{figure}

 \begin{figure*}
    \includegraphics[width=0.9\textwidth]{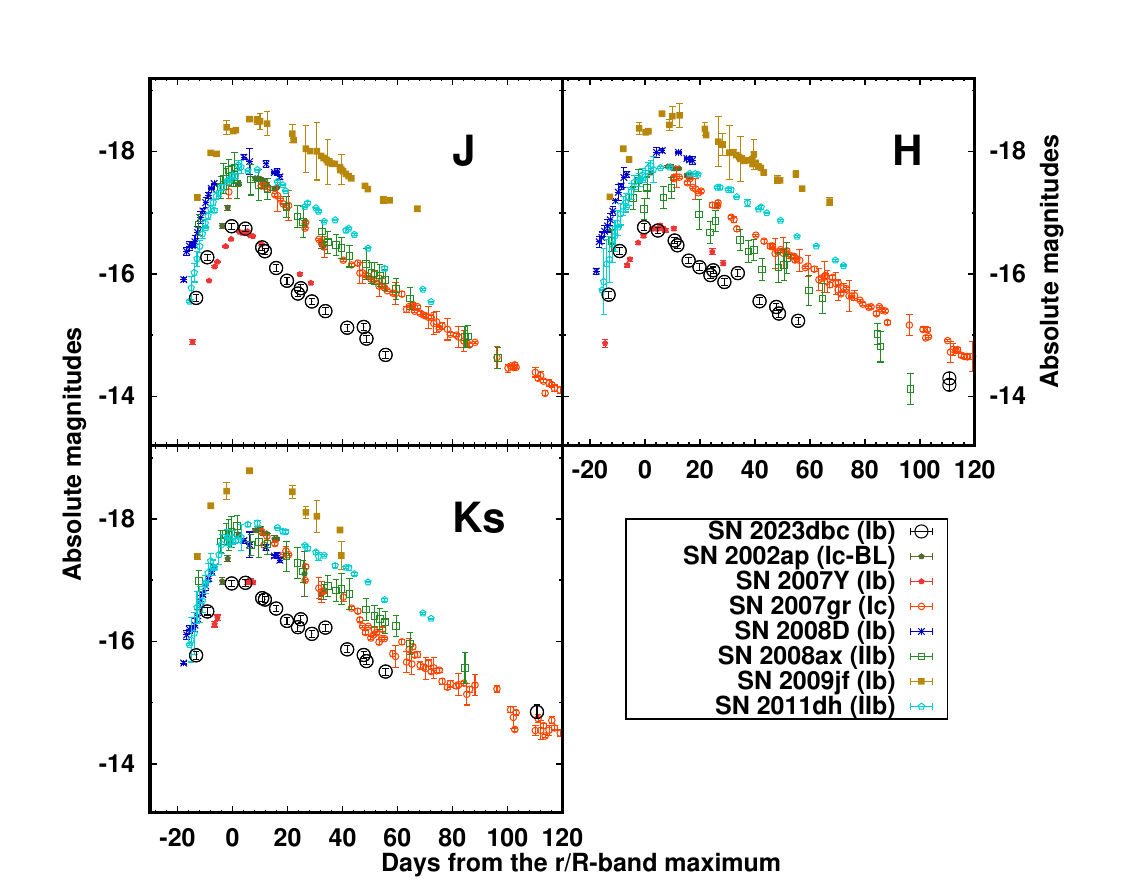} 
   \caption{NIR $J$, $H$, and $K_{\mathrm{s}}$-band absolute magnitude light curves of SN~2023dbc{\bf , compared with those of} other SNe~Ib/c and IIb. {\bf The light curves were corrected for extinction using} $A_J=1.0$, $A_H=0.7$, and $A_{K_{\mathrm{s}}}=0.4$\,mag{\bf , assuming a} distance modulus of $\mu=29.8$\,mag. {\bf For} SN~2023dbc, the {\bf reference epoch ($t=0$)} is the date of {\bf the} $r$-band maximum, {\bf whereas for the comparison objects, the reference epochs are as defined in their} respective {\bf literature}. Alt text: Three panels showing absolute magnitude evolution. The horizontal axis represents the phase in days relative to the maximum light.}
\end{figure*}

 \section{Observations and Data reduction}
NIR $JHK_{\mathrm{s}}$-band photometric observations were {\bf obtained over} 17 nights between 2023 March 15.6 and July 17.5 using the kSIRIUS instrument {\bf mounted at the Cassegrain focus of} the 1-m telescope at Kagoshima University \citep{Nagayama2024}. Optical $griz$-band observations {\bf were collected on} 7 nights from {\bf March 19.6 to April 13.9} using the {\bf GROWTH-India} Telescope (GIT; \cite{Kumar2022}). In addition, publicly available optical $g$- and $r$-band data from the Zwicky Transient Facility (ZTF; \cite{Bellm2019}) {\bf and} $o$-band data from {\bf the Asteroid Terrestrial-impact Last Alert System (ATLAS; \cite{Tonry2018})} were {\bf integrated into the analysis}.
 
 Data reduction {\bf for the} kSIRIUS and GIT imaging data was performed {\bf through a standard procedure}. After {\bf initial preprocessing}, we performed point-spread function (PSF) fitting photometry on all {\bf images}. Photometric calibration for the SN and {\bf reference stars} was {\bf calibrated against} the Pan-STARRS \citep{Chambers2016} and 2MASS \citep{Cutri2003} catalogues for the optical and NIR data, respectively. The {\bf resulting} multi-band light curves are {\bf presented} in Figure~1.

Spectroscopic {\bf observation} was {\bf conducted} on 2023 March 24 using the Himalayan Faint Object Spectrograph and Camera (HFOSC), {\bf mounted on} the 2-m Himalayan Chandra Telescope (HCT) at the Indian Astronomical Observatory (IAO), Hanle, India. {\bf A} slit width of $1\arcsec.92$ {\bf and the Grism~7 (gr7) were employed for the observation}. A {\bf blue} spectrophotometric standard star was also {\bf observed} on the same night {\bf for flux calibration}. Data reduction followed standard {\bf IRAF-based} procedures {\bf (or other software if applicable)}. Wavelength calibration was performed using a FeAr {\bf arc} lamp, and flux calibration was {\bf achieved using the aforementioned standard star}. {\bf We also incorporated an} additional spectrum {\bf retrieved} from the TNS public database. The wavelengths of these spectra were corrected to the rest frame {\bf adopting} a redshift of $z=0.0023$ \citep{Springob2005}. To determine the He~{\sc i} line velocity, we measured the absorption minimum by fitting a Gaussian function{\bf , where} the measurement error was {\bf estimated to be} $\sim 210~\mathrm{km~s^{-1}}$.

 \section{Results}
\subsection{Distance and extinction}

The host galaxy M108 is a {\bf nearby system located at a low redshift}. {\bf Given the large scatter among the distance estimates derived from various methods and the lack of Cepheid-based calibration, we adopted a distance of $9.9 \pm 0.7$\,Mpc for the host galaxy. This value is based on the average of Tully--Fisher relation measurements reported in the literature} \citep{dis1,dis2,dis3,dis4,dis5,dis6,dis7,dis8,dis9,dis10,dis11,dis12,dis13,dis14,dis15,dis16,dis17}.

{\bf First, we accounted for Galactic extinction ($A_V = 0.05$\,mag) using the dust extinction maps of \citet{Schlafly2011}. Regarding the host-galaxy extinction, the equivalent-width method using Na~{\sc i}~D absorption lines could not be applied due to the low signal-to-noise (S/N) ratio of our optical spectra. Therefore, we relied on a template color-evolution comparison for the extinction correction. 
The $r-J$, $r-H$, and $r-K_{\mathrm{s}}$ color evolutions were compared with those of other Type~Ib, Ic, and IIb SNe (Figure~2), where the comparison samples were corrected for extinction according to their respective references. We found that the color evolutions of SN~2023dbc are remarkably consistent with those of the Type~IIb SN~2011dh, as well as the broad-lined Type~Ic SN~2002ap.}

{\bf In contrast, the $r-K_{\mathrm{s}}$ and $r-H$ color evolutions of the other SNe in our sample are relatively constant. Based on these comparisons, we adopted the color evolutions of SNe~2011dh and 2002ap as templates for our extinction estimation. We determined the color excesses to be $E(r-J) = 2.27 \pm 0.12$\,mag, $E(r-H) = 2.56 \pm 0.08$\,mag, and $E(r-K_{\mathrm{s}}) = 2.94 \pm 0.07$\,mag. By assuming the standard Galactic extinction law with $R_V = 3.1$ \citep{CCM1989}, these values yield a total visual extinction of $A_V = 4.07 \pm 0.07$\,mag.}

{\bf Using a similar approach, we also estimated the extinction using $i-J$, $i-H$, and $i-K_{\mathrm{s}}$ color templates, which yielded color excesses of $E(i-J) = 1.51 \pm 0.09$\,mag, $E(i-H) = 1.85 \pm 0.06$\,mag, and $E(i-K_{\mathrm{s}}) = 2.15 \pm 0.14$\,mag. These values lead to $A_V = 4.05 \pm 0.09$\,mag, which is highly consistent with the results derived from the $r-JHK_{\mathrm{s}}$ templates. Therefore, we adopt $A_V = 4.1 \pm 0.1$\,mag as the host-galaxy extinction throughout this paper.}

{\bf Alternatively, we cross-checked the extinction value using the optical spectrum. First, we performed template fitting with GELATO \citep{Harutyunyan2008} while varying the reddening as a free parameter. The fit quality remained high for $E(B-V)$ in the range of 1.0--1.3\,mag; the latter value is consistent with the extinction derived from our color-template analysis ($A_V = 4.1$\,mag). To further confirm both the classification and reddening, we employed SNID-SAGE \citep{Stoppa2026}, which showed an excellent match with the broad-lined Type~Ic (SN~Ic-BL) PTF10vgv. These results demonstrate that the optical spectral features are consistent with those of an SN~Ic-BL, and the overall spectral slope is fully compatible with the high extinction ($A_V = 4.1$\,mag) determined above.}
    
 \subsection{Light curves}
 The $g, r, o, J, H$, and $K_{\mathrm{s}}$-band light curves are shown in Figure~1. We performed polynomial {\bf fitting} to the $r, o, J, H$, and $K_{\mathrm{s}}$-band light curves to estimate {\bf their} peak magnitudes and {\bf epochs} of maximum light. {\bf The resulting} maximum magnitudes {\bf are} 16.8, 16.5, 13.9, 13.6, and 13.2\,mag, {\bf which were reached} on MJD 60032.8, 60030.6, 60031.2, 60031.1, and 60031.1, respectively. {\bf We define the epoch of} the $r$-band maximum as $t = 0$ throughout this paper.

{\bf To constrain the explosion epoch}, we performed a fit to the {\bf rising phase of the} $r$ and $o$-band light curves using a modified exponential {\bf function}. {\bf The fit yielded an estimated} explosion date {\bf of} MJD~60017.6. Independently, we {\bf constrained} the explosion date {\bf to} MJD~$60017.85 \pm 0.75$ (or simply MJD~60017.85) {\bf based on the midpoint} between the {\bf first} discovery and the last non-detection {\bf epochs}. {\bf This result} is {\bf highly} consistent with the {\bf value} obtained from the {\bf light-curve fitting}. Therefore, we {\bf adopt} a rise time of 14.9\,days throughout this paper.

We obtained the $rJHK_{\mathrm{s}}$-band absolute magnitudes of SN~2023dbc by correcting for the Galactic \citep{Schlafly2011} and host-galaxy extinctions ($A_V = 4.1$; see \S~3.1), {\bf assuming a} distance modulus of $\mu = 29.8$. The peak absolute magnitudes {\bf were determined to be} $M_r = -16.3 \pm 0.2$\,mag, $M_J = -16.8 \pm 0.2$\,mag, $M_H = -16.8 \pm 0.2$\,mag, and $M_{K_{\mathrm{s}}} = -17.0 \pm 0.2$\,mag. The $J, H$, and $K_{\mathrm{s}}$-band light curves are compared with those of other SNe~Ib, Ic, and Ic-BL in Figure~3. The absolute magnitudes {\bf of SN~2023dbc lie} at the {\bf faint} end of the {\bf distribution for} these SNe, but are {\bf broadly} comparable to those of typical SNe~Ib/c. After {\bf the} maximum, the decline rates of SN~2023dbc are consistent with those of the {\bf comparison}.

{\bf SN~2007Y is another example of an intrinsically faint Type~Ib SN}, {\bf with a} peak $V$-band absolute magnitude {\bf of} $M_V = -16.5$\,mag \citep{Stritzinger2009}. Considering the {\bf uncertainties in both} extinction and distance, {\bf its} luminosity is comparable to {\bf that of} SN~2023dbc. However, as {\bf will be} discussed in \S~3.4, {\bf their} spectral features {\bf exhibit significant differences}.

 \subsection{Bolometric light curves}

 \begin{figure}
    \includegraphics[width=0.45\textwidth]{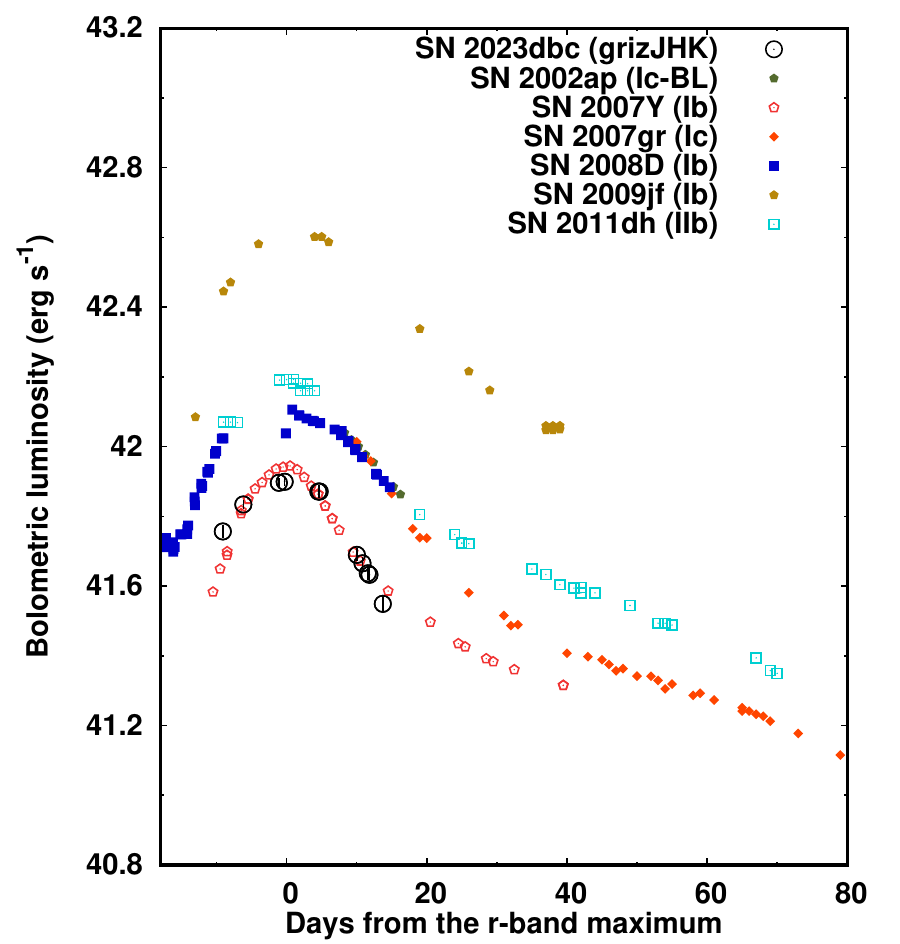} 
   \caption{Quasi-bolometric light curve of SN~2023dbc, {\bf derived by integrating} the $g, r, i, z, J, H$, and $K_{\mathrm{s}}$-band photometric data. {\bf For comparison, the bolometric light curves of other SNe are also shown, including} SNe~2002ap \citep{Foley2003,Yoshii2003,Tomita2006}, 2007Y \citep{Stritzinger2009}, 2007gr \citep{Valenti2008b,Hunter2009}, 2008D \citep{Modjaz2009}, 2009jf \citep{Valenti2011}, and 2011dh \citep{Ergon2014}{\bf , which were constructed from $BVRIJHK_{\mathrm{s}}$-band data}. 
   Alt text: The vertical and horizontal axes {\bf represent the luminosity in $\mathrm{erg\,s^{-1}}$ and the phase in days relative to} maximum light, respectively.}
\end{figure}

The quasi-bolometric light curve was constructed by integrating the {\bf flux densities at the effective wavelengths of each filter} \citep{Fukugita1996}, {\bf assuming linear interpolation between the data points}. We utilized the $g, r, i, z, J, H$, and $K_{\mathrm{s}}$-band {\bf fluxes}, which were {\bf derived from the observed magnitudes after correcting for} the extinction of $A_V = 4.1$\,mag (see \S~3.1) and {\bf adopting} the distance to the host galaxy M108. {\bf Following} \citet{Lyman2014}, we assumed that the $grizJHK_{\mathrm{s}}$ {\bf bands cover approximately 80\% of} the total {\bf bolometric} flux. The {\bf resulting} quasi-bolometric light curve is shown in Figure~4. The peak {\bf bolometric} luminosity was {\bf estimated to be} $\sim 5.8 \times 10^{41}\,\mathrm{erg\,s^{-1}}$.

The quasi-bolometric light curves {\bf were compared with other stripped-envelope SNe}, including SNe~2002ap \citep{Foley2003,Yoshii2003,Tomita2006}, 2007Y \citep{Stritzinger2009}, 2007gr \citep{Valenti2008b,Hunter2009}, 2008D \citep{Modjaz2009}, 2009jf \citep{Valenti2011,Sahu2011}, and 2011dh \citep{Ergon2014}. {\bf For these objects, we accounted for distance and extinction using values reported in the literature}. The fraction of the $BVRIJHK_{\mathrm{s}}$-band flux to the total bolometric flux {\bf for this comparison sample was also assumed to be $\sim 80\%$, the same as that used for SN~2023dbc.}

The peak luminosity {\bf of SN~2023dbc was significantly lower} than {\bf those} of other SNe~Ib, Ic, and Ic-BL. SN~2007Y is {\bf recognized as} one of the faintest Type~Ib SNe {\bf within the sample studied by} \citet{Lyman2016}. {\bf Notably, even after accounting for the uncertainties in distance and extinction, the luminosity of SN~2023dbc remains remarkably low, appearing even fainter than that of SN~2007Y.}
  
The peak luminosity of stripped-envelope SNe is {\bf primarily} determined by the radioactive $^{56}$Ni mass and the rise time \citep{Arnett1982}. Using a rise time of $14.9$\,days and the peak luminosity {\bf derived above}, the radioactive $^{56}$Ni mass was estimated to be $(3.8 \pm 0.1) \times 10^{-2}\,M_{\odot}$. This mass {\bf lies at} the lower edge of the distribution {\bf for} SNe~Ib/c and Ic-BL \citep{Lyman2016, Afsariardchi2021}{\bf . Specifically, it} is smaller than the $^{56}$Ni masses {\bf reported for the SN~Ic-BL~2002ap ($0.07\,M_{\odot}$; \cite{Mazzali2002}) and the SN~Ib~2007Y ($0.06\,M_{\odot}$; \cite{Stritzinger2009})}.

 \subsection{Spectra}

\begin{figure}
    \includegraphics[width=0.45\textwidth]{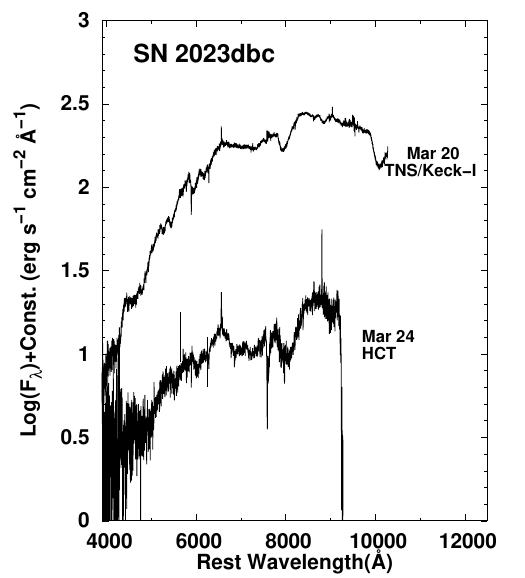} 
  \caption{{\bf Optical spectra} of SN~2023dbc obtained at $t = -8$\,d and $t = -4$\,d {\bf relative} to the $r$-band maximum. The spectra are shown {\bf without correction} for extinction. Alt text: The vertical axis represents the flux density in units of $\mathrm{erg\,s^{-1}\,cm^{-2}\, {\AA}^{-1}}$ on a logarithmic scale, and the horizontal axis represents the observed wavelength in {\AA}.}
\end{figure}

\begin{figure}
    \includegraphics[width=0.45\textwidth]{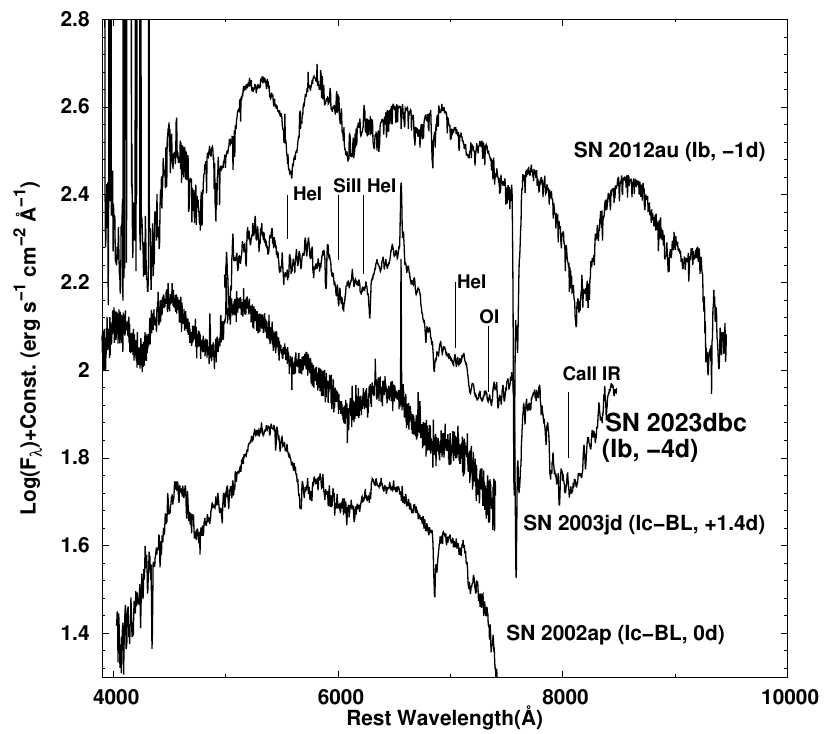} 
 \caption{{\bf Extinction-corrected spectrum} of SN~2023dbc ($A_V = 4.1$\,mag 
 and $R_V = 3.1$). {\bf The spectrum was smoothed using a 5-pixel boxcar window
 to enhance the signal-to-noise ratio.} For comparison, {\bf spectra of the
 Type~Ic-BL~SNe~2002ap, 2003jd \citep{Valenti2008a}, and the Type~Ib~SN~2012au
 \citep{Takaki2013} are also shown}. The $t = 0$\,d spectrum of SN~2002ap was
 {\bf retrieved} from the Weizmann Interactive Supernova data Repository
 (WISeREP; \cite{Yaron2012}) 
 {\bf with original data
 reported in} \citet{Foley2003} and \citet{Kinugasa2002}. All wavelengths {\bf have been shifted} to the rest frame using the recession velocity of each host
 galaxy. Alt text: The vertical axis represents the flux density on a
 logarithmic scale, and the
 horizontal axis represents the rest-frame wavelength.}
\end{figure}

Figure~5 shows the spectral evolution of SN~2023dbc. {\bf Although} the observed spectra are significantly reddened {\bf owing} to the high extinction, several absorption features {\bf can be} clearly identified. {\bf Specifically}, the He~{\sc i} $\lambda$5876, $\lambda$6678, and $\lambda$7065 lines are present. The Si~{\sc ii}$\lambda$6355 absorption feature is also visible {\bf around} $t = 4$\,d. For the subsequent analysis, these spectra were corrected for extinction using $A_V = 4.1$\,mag, assuming the standard reddening law with $R_V = 3.1$.

In Figure~6, the extinction-corrected spectrum is compared with {\bf those of the well-studied Type~Ib~SN~2012au \citep{Takaki2013} and the SNe~Ic-BL~2002ap \citep{Kinugasa2002, Foley2003} and 2003jd \citep{Valenti2008a}}. Compared to SN~2012au, the absorption lines in SN~2023dbc appear {\bf broader and more blended}. The blueshifted O~{\sc i}~$\lambda$7774 absorption feature is clearly identified, indicating an expansion velocity higher than that of SN~2012au. Furthermore, the Ca~{\sc ii} NIR triplet {\bf exhibits a significantly higher velocity than that observed in SN~2012au.}

While the overall spectral profile resembles that of the Type~Ic-BL~SN~2003jd, the He~{\sc i}~$\lambda$5876, $\lambda$6678, and $\lambda$7065 absorption features are distinctly visible in SN~2023dbc. Notably, the absorption features of Si~{\sc ii}~$\lambda$6355 and 
He~{\sc i}~$\lambda$6678 are marginally resolved; in contrast, these features are typically blended in SNe~Ic-BL {\bf because of} their significantly higher expansion velocities.



\begin{figure}
    \includegraphics[width=0.45\textwidth]{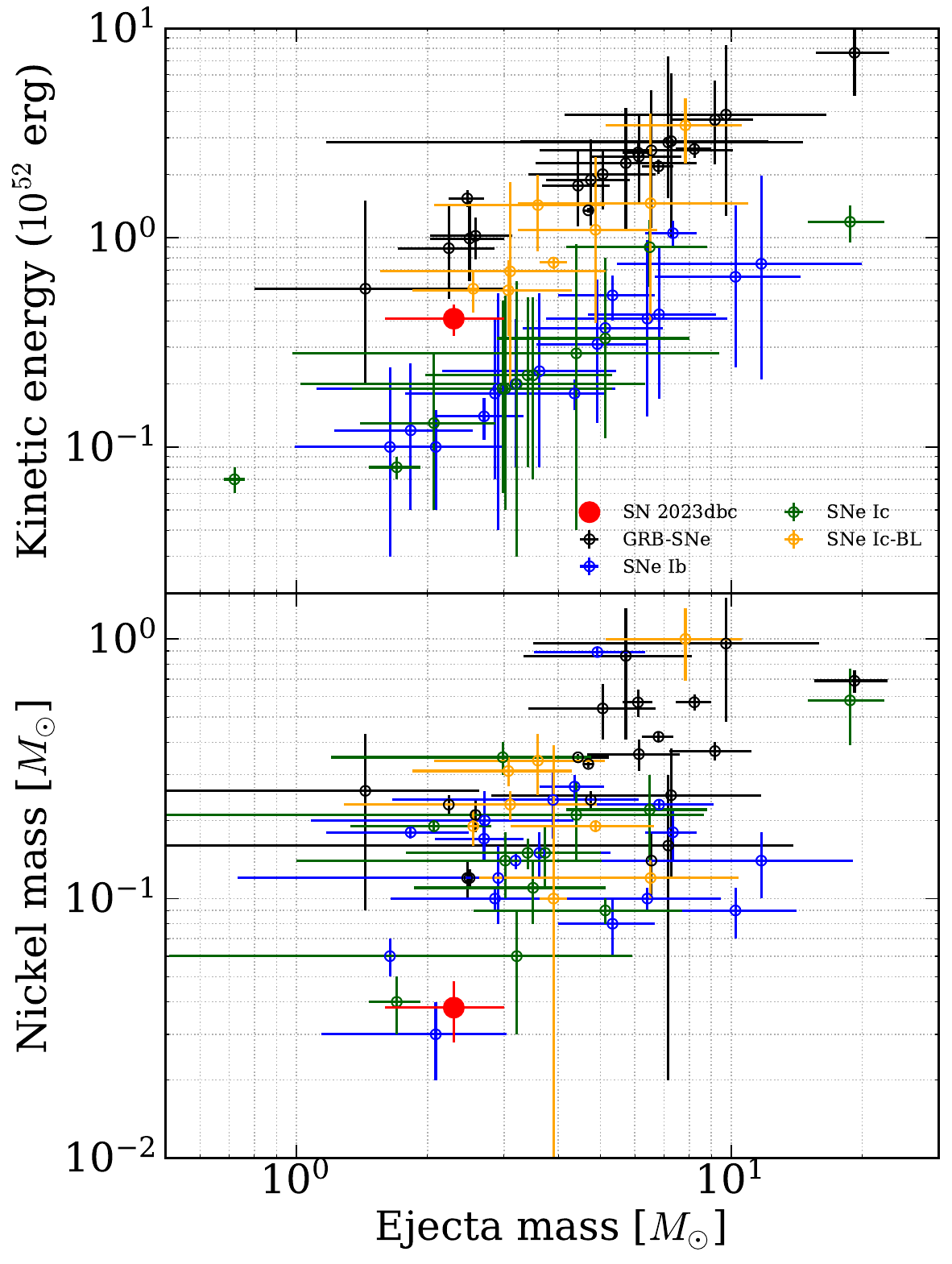} 
    \caption{{\bf (Upper panel) Ejecta mass ($M_{\mathrm{ej}}$) versus kinetic energy ($E_{\mathrm{kin}}$) for SN~2023dbc compared with the large sample from \citet{Cano2013}. The explosion properties of SN~2023dbc are consistent with those of typical SNe~Ib/c.} (Lower panel) {\bf Similar to the upper panel, but showing the correlation between} $M_{\mathrm{ej}}$ and the radioactive $^{56}\mathrm{Ni}$ mass ($M_{\mathrm{Ni}}$). {\bf In both panels, SN~2023dbc is located within the distribution of standard stripped-envelope SNe.}
    Alt text: The vertical axis shows the kinetic energy in logarithmic scale, in units of erg in upper panel, and the nickel mass 
    in logarithmic scale, in units of solar mass in lower panel. 
    The horizontal axis shows the ejecta mass in logarithmic scale, 
    in units of solar mass.}
\end{figure}

\section{Discussion}

The ejecta mass ($M_{\mathrm{ej}}$) and kinetic energy ($E_{\mathrm{kin}}$) are fundamentally related to the opacity ($\kappa$), the diffusion timescale ($\tau_{\mathrm{diff}}$), and the expansion velocity ($v$) of the supernova ejecta \citep{Arnett1982}. These relationships are governed by the following scaling laws:
\begin{equation}
\tau_{\mathrm{diff}} \propto \kappa^{1/2} M_{\mathrm{ej}}^{3/4} E_{\mathrm{kin}}^{-1/4},
\end{equation}
\begin{equation}
v \propto M_{\mathrm{ej}}^{-1/2} E_{\mathrm{kin}}^{1/2}.
\end{equation}
We adopt the observed rise time ($\tau_{\mathrm{rise}} = 14.9$\,days) as a proxy for the diffusion timescale. The expansion velocity was measured to be $v = 16,000\,\mathrm{km\,s^{-1}}$ using the spectrum at $t = -4$\,d; however, the line velocities exhibit a rapid decline during this phase. Assuming a decline rate of 500 km~$s^{-1}$ d$^{-1}$ (e.g., \cite{Sahu2011}), the velocity at $t = 0$\,d is inferred to be $v \approx 14,000\,\mathrm{km\,s^{-1}}$. We use this value to estimate the physical properties of the ejecta.

Since the absolute value of $\kappa$ is model-dependent, we employ a relative scaling method to mitigate this uncertainty, using SN~2012au as a reference. For SN~2012au, \citet{Takaki2013} reported a He~{\sc i} line velocity of $13,500\,\mathrm{km\,s^{-1}}$ and a rise time of $16.3$\,days, corresponding to $M_{\mathrm{ej}} = 6.5\,M_{\odot}$ and $E_{\mathrm{kin}} = 5.1 \times 10^{51}$\,erg. By applying the scaling laws, we obtain the explosion properties for SN~2023dbc as $M_{\mathrm{ej}} = 7.1 \pm 1.4\,M_{\odot}$ and $E_{\mathrm{kin}} = (5.7 \pm 3.2) \times 10^{51}$\,erg.

{\bf We also independently estimated the ejecta mass as $M_{\mathrm{ej}} = 1.6 \pm 0.3\,M_{\odot}$ and the kinetic energy as $E_{\mathrm{kin}} = (1.6 \pm 0.9) \times 10^{51}\,\mathrm{erg}$, based on the scaling-law method using the reference values for the energetic Type~Ic-BL~SN~2002ap from \citet{Lyman2016}. For this calibration, we adopted a reference velocity of $13,000\,\mathrm{km\,s^{-1}}$ and a rise time of $12.45$\,days for SN~2002ap. These results are notably lower than our initial estimates derived from SN~2012au. This discrepancy likely arises from differences in the assumed model parameters and coefficients, such as the optical opacity ($\kappa_{\mathrm{opt}}$), adopted in different studies. By combining these two independent calibrations, we aimed to account for the systematic uncertainties arising from different model assumptions, such as the opacity and ejecta structure inherent in each study. Consequently, we obtained $M_{\mathrm{ej}} = {\bf 4.4 \pm 2.8}\,M_{\odot}$ and $E_{\mathrm{kin}} = {\bf (3.7 \pm 2.1)} \times 10^{51}$\,erg as our final estimates for SN~2023dbc.}
 
\citet{Cano2013} also presented an analytic investigation of a collected sample of Type~Ib, Ic, Ic-BL, and GRB-SNe. {\bf Figure~7 shows the kinetic energy plotted against the ejecta mass for SN~2023dbc, along with the sample from \citet{Cano2013}. Our derived explosion properties are consistent with the distribution of SNe~Ic-BL, although both parameters are located toward the lower end of the population.} This finding is consistent with the overall spectral profile, which exhibits {\bf the} broad features characteristic of {\bf moderately} energetic events.
  
\begin{figure}
    \includegraphics[width=0.45\textwidth]{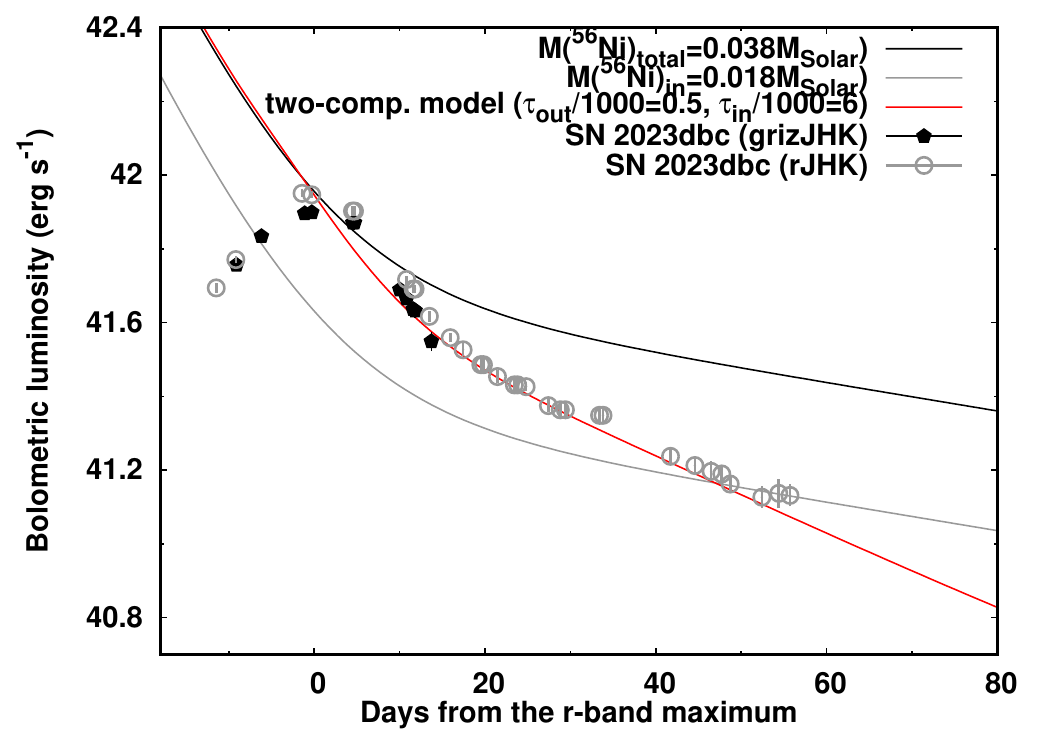} 
    \caption{Quasi-bolometric light curve of SN~2023dbc. In addition to the {\bf flux} integrated over the $grizJHK_{\mathrm{s}}$ bands, the curve derived from {\bf the integration of} $rJHK_{\mathrm{s}}$ {\bf data} is also shown {\bf for comparison}. The data are {\bf fitted} with two-component model curves \citep{Maeda2003} {\bf characterized by} $\gamma$-ray optical depths of $\tau_{\mathrm{out}}/1000 = 0.5$ and $\tau_{\mathrm{in}}/1000 = 6$. The model assumes an {\bf outer-layer} $^{56}\mathrm{Ni}$ mass of $M(^{56}\mathrm{Ni})_{\mathrm{out}} = 0.02\,M_{\odot}$ and an {\bf inner-layer} mass of $M(^{56}\mathrm{Ni})_{\mathrm{in}} = 0.018\,M_{\odot}$. The total synthesized $^{56}\mathrm{Ni}$ mass {\bf adopted} for this two-component model is $M(^{56}\mathrm{Ni})_{\mathrm{total}} = (3.8 \pm 0.1) \times 10^{-2}\,M_{\odot}$.   
    Alt text: The vertical axis is scaled in erg per second.
    The horizontal axis is scaled in days from the maximum light.
    }
\end{figure}


\citet{Mazzali2007} discussed the possibility of {\bf an aspherical explosion scenario based on a} theoretical analysis of the line {\bf profiles in} the nebular-phase spectra {\bf of} the faint Type~Ic-BL~SN~2002ap. {\bf To investigate this further, we calculated two-component} model light curves {\bf following the methodology described in} \citet{Maeda2003}. In this model, the ejecta structure is assumed to {\bf consist} of an {\bf outer layer with low density and a high-density inner core}. The model {\bf parameters are} described as follows:

\begin{align}
L_{\mathrm{bol}} &= M(^{56}\mathrm{Ni})_{\mathrm{in}}e^{-t_{\mathrm{d}}/113}[\epsilon_{\gamma}(1-e^{-\tau_{\mathrm{in}}})+\epsilon_{e^{+}}] \nonumber \\
& \quad + M(^{56}\mathrm{Ni})_{\mathrm{out}}e^{-t_{\mathrm{d}}/113}[\epsilon_{\gamma}(1-e^{-\tau_{\mathrm{out}}})+\epsilon_{e^{+}}]
\end{align}

To demonstrate the {\bf applicability of this} model, we constructed the quasi-bolometric light curve of SN~2023dbc by integrating the flux using only the $r, J, H,$ and $K_{\mathrm{s}}$-band data {\bf up to} $t \approx 60$\,d, {\bf following} the method described in \S~3. {\bf For this specific calculation}, the ratio of the $rJHK_{\mathrm{s}}$ {\bf integrated flux} to the total bolometric flux is assumed to be {\bf $\sim 80\%$}. We {\bf performed a fit of} the model light curve to the observations (see Figure~8). {\bf To reproduce the observed light curve, the outer-layer ejecta require a $\gamma$-ray optical depth of $\tau_{\mathrm{out}}/1000 \approx 0.5$, while the inner core requires $\tau_{\mathrm{in}}/1000 \approx 6$. The significant ratio of $\tau_{\mathrm{in}}/\tau_{\mathrm{out}} = 12$} is indicative of a {\bf steep density gradient} between the outer ejecta and the {\bf high-density} core. This trend is consistent with {\bf the results reported} for the highly energetic SNe~1998bw and 2002ap \citep{Maeda2003}.

Considering the similarity with these SNe, an aspherical {\bf ejecta structure is strongly suggested}. In this asymmetric {\bf configuration}, relatively high-velocity ejecta are expected to be rich in $^{56}\mathrm{Ni}$ along the jet {\bf axis}, {\bf while} a ring-like geometry of oxygen-rich gas is expected {\bf in the equatorial plane} \citep{Maeda2008}. If the viewing angle is {\bf off-axis (e.g., perpendicular to the jet)}, a {\bf more moderate} energy would be {\bf observed} compared to the {\bf on-axis} direction. This scenario is consistent with the picture of a bipolar explosion, albeit less extreme than {\bf prominent} cases {\bf such as} SN~2008D \citep{Tanaka2009b}.


{\bf Alternatively, we consider a fallback scenario} in which the inner part of the ejecta {\bf accretes} onto the central remnant, such as a black hole. In this case, the newly synthesized elements in the inner region, including {\bf a significant fraction of} the radioactive $^{56}\mathrm{Ni}$, would {\bf be sequestered into} the central {\bf object}. Another possibility is a low-energy explosion, resulting in a {\bf very} small synthesized $^{56}\mathrm{Ni}$ mass {\bf (e.g., $< 0.01\,M_{\odot}$)}, as seen in extremely faint SNe~IIP \citep{Turatto1998, Zampieri2003}. {\bf However}, the radioactive $^{56}\mathrm{Ni}$ mass of SN~2023dbc is $\approx 3.8 \times 10^{-2}\,M_{\odot}$, which is larger than {\bf typical values} predicted by {\bf simple} fallback scenarios. In {\bf such} scenarios, the ejecta mass is {\bf expected to be large} while the total explosion energy {\bf remains} small, {\bf which contrasts with} the moderately large kinetic energy and ejecta mass {\bf derived} for SN~2023dbc. {\bf These} arguments {\bf suggest that a} partial fallback of the inner core may {\bf have occurred}. The {\bf inferred} weak asphericity of the ejecta structure supports {\bf this} fallback mechanism{\bf , where} the inner core is relatively dense, but not significantly denser than {\bf those of} SNe~1998bw and 2002ap. Based on these {\bf considerations}, an aspherical explosion scenario {\bf involving} the formation of a central black hole is consistent with the observed properties of SN~2023dbc, {\bf although} further analysis of {\bf late-phase} data {\bf is} needed {\bf to confirm this picture}.

The ratio of {\bf $M(^{56}\mathrm{Ni})_{\mathrm{in}}/M(^{56}\mathrm{Ni})_{\mathrm{out}} \approx 0.9$} is larger than {\bf those observed in} highly energetic SNe {\bf such} as SNe~1998bw and 2002ap \citep{Maeda2003}. This is consistent with {\bf a degree of moderate} ejecta mixing. {\bf Such moderate mixing}, often linked to an asymmetric explosion, may contribute to the formation of He~{\sc i} absorption lines \citep{Cano2014}. The absorption features of He~{\sc i}~$\lambda$5876, $\lambda$6678, and $\lambda$7065 are {\bf typically produced} by non-thermal excitation {\bf driven by} the radioactive decay of $^{56}\mathrm{Ni}$ \citep{Hachinger2012}. {\bf While} SNe~Ic-BL {\bf occasionally} exhibit these He~{\sc i} lines due to this mechanism, SN~2023dbc appears to be {\bf a transitional} object between normal SNe~Ib and SNe~Ic-BL. Furthermore, the short rise time observed for SN~2023dbc is consistent with {\bf a significant fraction of} $^{56}\mathrm{Ni}$ being distributed in the outer {\bf layers} of the ejecta.

Based on the {\bf estimated} total ejecta mass of $M_{\mathrm{ej}} \approx 2.3\,M_{\odot}$ and assuming the $M_{\alpha}\text{--}M_{\mathrm{MS}}$ relation \citep{Sugimoto1980}, the main-sequence mass of the progenitor is estimated to be $M_{\mathrm{MS}} \approx 15\,M_{\odot}$. This initial mass is consistent with the progenitors of {\bf normal} SNe~Ib/c, such as SN~1994I \citep{Nomoto1994}, but is {\bf notably} smaller than the estimates for SNe~2008D \citep{Tanaka2009} and 2002ap \citep{Mazzali2002, Mazzali2007}. Generally, the hydrogen envelope of a massive star is {\bf effectively} stripped by radiation-driven winds {\bf for very massive stars} \citep{Castor1975}. {\bf However, for SN~2023dbc, whose} ejecta mass is comparable to typical SNe {\bf despite a relatively small} $^{56}\mathrm{Ni}$ mass, {\bf binary interaction may be required} to {\bf sufficiently} strip the envelope. {\bf Taken together}, these arguments suggest that SN~2023dbc {\bf originated} from an aspherical explosion with {\bf partial} fallback, {\bf resulting from a moderate-mass} progenitor in a binary system.

 
\section{Conclusion} 
SN~2023dbc is a highly reddened Type~Ib~SN, characterized by a relatively faint intrinsic peak {\bf luminosity}, a fast rise time, and {\bf high-velocity He{\sc i} absorption features}. Based on these observational properties, we {\bf estimated} the explosion parameters of this object. The faint peak and short rise time constrain the ejected $^{56}\mathrm{Ni}$ mass to a relatively low value of $M(^{56}\mathrm{Ni}) = (3.8 \pm 0.1) \times 10^{-2}\,M_{\odot}$. Furthermore, the high expansion velocity and short characteristic timescale yield {\bf an} ejecta mass of $M_{\mathrm{ej}} = 2.3 \pm 0.7\,M_{\odot}$ and a kinetic energy of $E_{\mathrm{kin}} = (4.1 \pm 0.7) \times 10^{51}$\,erg. The {\bf relatively small ratio of the synthesized $^{56}\mathrm{Ni}$ mass} to the ejecta mass is consistent with the explosion {\bf models proposed for} the intrinsically faint {\bf Type~Ic-BL} SN~2002ap.

The analysis {\bf using} the two-component model {\bf demonstrates} that the density profile is steep in the outer {\bf layers} while {\bf maintaining} a dense inner core. This structure {\bf suggests an aspherical explosion geometry}, {\bf albeit one that is less pronounced} than that of SN~2002ap. The short rise time {\bf further} supports {\bf the presence of} moderate mixing {\bf within} the ejecta. The {\bf relatively} low $^{56}\mathrm{Ni}$ mass is {\bf well-explained} by a partial fallback scenario, {\bf which} is consistent with the hypothesis that {\bf the SN was observed from a direction almost perpendicular to the high-velocity (jet-like) ejecta}. This {\bf configuration} is similar to {\bf the off-axis view proposed for the aspherical explosion structure of} SN~1998bw \citep{Maeda2008}.

The relatively {\bf small} ejecta mass indicates that the {\bf zero-age main-sequence mass} of the progenitor was likely $M_{\mathrm{ZAMS}} \approx 15\,M_{\odot}$. The progenitor {\bf was presumably} stripped of its hydrogen envelope {\bf via} binary interaction, {\bf leaving a helium star as the pre-supernova progenitor}. This suggests that SN~2023dbc resulted from an aspherical explosion {\bf associated with the} partial fallback {\bf of the inner core onto a} black hole, {\bf originating} from a moderately massive star in a close binary system.

\begin{ack}
 We are grateful to graduate and undergraduate students for performing 
 the near-infrared observations. This work was supported by Grant-in-Aid for Scientific Research (C) 22K03676. The Kagoshima University 1 m telescope is a member of the Optical and Infrared Synergetic Telescopes for Education and Research (OISTER) program funded by the MEXT of Japan. This work was supported by JSPS Bilateral Program Number JPJSBP 120227709. 
 The GROWTH-India Telescope (GIT) is a 70-cm telescope
with a 0$^{\circ}$. 7 field of view, set up by the 
 Indian Institute of
 Astrophysics (IIA) and the Indian Institute of Technology
Bombay (IITB) with funding from Indo-US Science and
Technology Forum and the Science and Engineering Research
Board, Department of Science and Technology, Government of
India. It is located at the Indian Astronomical Observatory
(IAO, Hanle). We acknowledge funding by the IITB alumni
batch of 1994, which partially supports the operation of the
telescope. Telescope technical details are available at
GROWTH-India website https://sites.google.com/view/growthindia/.
\end{ack}

\begin{appendix}

\begin{table*}
   \caption{Log of NIR photometry of SN 2023dbc obtained with kSIRIUS.}
    \centering
    \begin{tabular}{lllllll}
      \hline
      MJD & $J$ & err($J$) & H & err($H$) & $K_{\mathrm s}$ & err($K_{\mathrm s}$) \\
      \hline
    60018.6 & 15.209 & 0.011 & 14.795 & 0.022 & 14.463 & 0.030 \\ 
60022.6 & 14.546 & 0.012 & 14.083 & 0.028 & 13.746 & 0.025 \\ 
60031.5 & 14.039 & 0.007 & 13.692 & 0.017 & 13.288 & 0.013 \\ 
60036.5 & 14.077 & 0.006 & 13.747 & 0.012 & 13.279 & 0.008 \\ 
60042.6 & 14.384 & 0.008 & 13.911 & 0.019 & 13.528 & 0.011 \\ 
60043.6 & 14.447 & 0.009 & 13.988 & 0.026 & 13.552 & 0.009 \\ 
60047.7 & 14.721 & 0.009 & 14.239 & 0.015 & 13.698 & 0.013 \\ 
60051.6 & 14.929 & 0.013 & 14.346 & 0.014 & 13.901 & 0.018 \\ 
60055.6 & 15.137 & 0.018 & 14.476 & 0.026 & 14.003 & 0.039 \\ 
60056.6 & 15.050 & 0.015 & 14.398 & 0.018 & 13.873 & 0.019 \\ 
60060.6 & 15.266 & 0.012 & 14.588 & 0.023 & 14.112 & 0.015 \\ 
60065.6 & 15.425 & 0.016 & 14.444 & 0.022 & 14.010 & 0.021 \\ 
60073.5 & 15.696 & 0.015 & 14.902 & 0.016 & 14.365 & 0.026 \\ 
60079.5 & 15.684 & 0.031 & 14.997 & 0.030 & 14.458 & 0.033 \\ 
60080.5 & 15.874 & 0.020 & 15.104 & 0.025 & 14.557 & 0.026 \\ 
60087.5 & 16.139 & 0.022 & 15.223 & 0.025 & 14.727 & 0.025 \\ 
      \hline
      \end{tabular}
\end{table*}

\begin{table*}
 \caption{Log of optical photometry of SN 2023dbc obtained with ZTF and GIT.}
     \centering
    \begin{tabular}{llllllllll}
    \hline
    MJD & $g$ & err($g$) & $r$ & err($r$) & $i$ & err($i$) & $z$ & $err(z)$ & Telescope \\
\hline
 60020.3 & 19.263 & 0.103 & 17.358 & 0.039 & $\cdot\cdot\cdot$ & $\cdot\cdot\cdot$ & $\cdot\cdot\cdot$ & $\cdot\cdot\cdot$ & ZTF \\ 
 60022.7 & $\cdot\cdot\cdot$ & $\cdot\cdot\cdot$ & 17.104 & 0.025 & 17.104 & 0.025 & 15.723 & 0.080 & GIT\\ 
 60025.6 & 18.750 & 0.051 & $\cdot\cdot\cdot$ & $\cdot\cdot\cdot$ & $\cdot\cdot\cdot$ & $\cdot\cdot\cdot$ & 15.528 & 0.048 & GIT \\ 
 60030.7 & 18.688 & 0.043 & $\cdot\cdot\cdot$ & $\cdot\cdot\cdot$ & 16.915 & 0.034 & 15.348 & 0.064 & GIT \\ 
 60036.3 & 19.017 & 0.114 & 16.931 & 0.047 & $\cdot\cdot\cdot$ & $\cdot\cdot\cdot$ & $\cdot\cdot\cdot$ & $\cdot\cdot\cdot$ & ZTF\\ 
 60041.8 & $\cdot\cdot\cdot$ & $\cdot\cdot\cdot$ & 17.483 & 0.050 & 17.483 & 0.050 & 15.757 & 0.059 & GIT \\ 
 60043.4 & $\cdot\cdot\cdot$ & $\cdot\cdot\cdot$ & 17.541 & 0.041 & $\cdot\cdot\cdot$ & $\cdot\cdot\cdot$ & $\cdot\cdot\cdot$ & $\cdot\cdot\cdot$ & ZTF \\ 
 60045.5 & 20.507 & 0.231 & 17.767 & 0.054 & $\cdot\cdot\cdot$ & $\cdot\cdot\cdot$ & 15.955 & 0.083 & GIT \\ 
 60047.9 & $\cdot\cdot\cdot$ & $\cdot\cdot\cdot$ & 17.425 & 0.040 & 17.425 & 0.040 & 16.114 & 0.096 & GIT \\ 
 60049.2 & $\cdot\cdot\cdot$ & $\cdot\cdot\cdot$ & 17.995 & 0.067 & $\cdot\cdot\cdot$ & $\cdot\cdot\cdot$ & $\cdot\cdot\cdot$ & $\cdot\cdot\cdot$ & ZTF \\ 
 60051.3 & $\cdot\cdot\cdot$ & $\cdot\cdot\cdot$ & 18.091 & 0.070 & $\cdot\cdot\cdot$ & $\cdot\cdot\cdot$ & $\cdot\cdot\cdot$ & $\cdot\cdot\cdot$ & ZTF \\ 
 60053.2 & $\cdot\cdot\cdot$ & $\cdot\cdot\cdot$ & 18.175 & 0.067 & $\cdot\cdot\cdot$ & $\cdot\cdot\cdot$ & $\cdot\cdot\cdot$ & $\cdot\cdot\cdot$ & ZTF \\ 
 60055.2 & $\cdot\cdot\cdot$ & $\cdot\cdot\cdot$ & 18.206 & 0.063 & $\cdot\cdot\cdot$ & $\cdot\cdot\cdot$ & $\cdot\cdot\cdot$ & $\cdot\cdot\cdot$ & ZTF \\ 
 60059.2 & $\cdot\cdot\cdot$ & $\cdot\cdot\cdot$ & 18.395 & 0.064 & $\cdot\cdot\cdot$ & $\cdot\cdot\cdot$ & $\cdot\cdot\cdot$ & $\cdot\cdot\cdot$ & ZTF\\ 
 60061.2 & $\cdot\cdot\cdot$ & $\cdot\cdot\cdot$ & 18.398 & 0.071 & $\cdot\cdot\cdot$ & $\cdot\cdot\cdot$ & $\cdot\cdot\cdot$ & $\cdot\cdot\cdot$ & ZTF \\  
 60065.2 & $\cdot\cdot\cdot$ & $\cdot\cdot\cdot$ & 18.432 & 0.071 & $\cdot\cdot\cdot$ & $\cdot\cdot\cdot$ & $\cdot\cdot\cdot$ & $\cdot\cdot\cdot$ & ZTF \\ 
 60076.3 & $\cdot\cdot\cdot$ & $\cdot\cdot\cdot$ & 18.769 & 0.090 & $\cdot\cdot\cdot$ & $\cdot\cdot\cdot$ & $\cdot\cdot\cdot$ & $\cdot\cdot\cdot$ & ZTF \\ 
 60078.3 & $\cdot\cdot\cdot$ & $\cdot\cdot\cdot$ & 18.823 & 0.097 & $\cdot\cdot\cdot$ & $\cdot\cdot\cdot$ & $\cdot\cdot\cdot$ & $\cdot\cdot\cdot$ & ZTF \\ 
 60084.2 & $\cdot\cdot\cdot$ & $\cdot\cdot\cdot$ & 18.948 & 0.105 & $\cdot\cdot\cdot$ & $\cdot\cdot\cdot$ & $\cdot\cdot\cdot$ & $\cdot\cdot\cdot$ & ZTF \\ 
 60086.2 & $\cdot\cdot\cdot$ & $\cdot\cdot\cdot$ & 18.884 & 0.137 & $\cdot\cdot\cdot$ & $\cdot\cdot\cdot$ & $\cdot\cdot\cdot$ & $\cdot\cdot\cdot$ & ZTF \\ 
 60090.2 & $\cdot\cdot\cdot$ & $\cdot\cdot\cdot$ & 18.896 & 0.110 & $\cdot\cdot\cdot$ & $\cdot\cdot\cdot$ & $\cdot\cdot\cdot$ & $\cdot\cdot\cdot$ & ZTF \\ 
 60092.3 & $\cdot\cdot\cdot$ & $\cdot\cdot\cdot$ & 19.283 & 0.111 & $\cdot\cdot\cdot$ & $\cdot\cdot\cdot$ & $\cdot\cdot\cdot$ & $\cdot\cdot\cdot$ & ZTF \\ 
 60105.3 & $\cdot\cdot\cdot$ & $\cdot\cdot\cdot$ & 19.447 & 0.129 & $\cdot\cdot\cdot$ & $\cdot\cdot\cdot$ & $\cdot\cdot\cdot$ & $\cdot\cdot\cdot$ & ZTF\\ 
      \hline
      \end{tabular}
\end{table*}

\end{appendix}

\bibliography{addsample}{}
\bibliographystyle{apj}

\end{document}